\documentclass[prl,aps,twocolumn]{revtex4}

\usepackage{graphicx,epsfig}
\usepackage{float}
\usepackage{dcolumn}
\usepackage{bm}

\begin{document}


\title{Current-Induced Resonant Motion of a Magnetic Vortex Core: Effect of Nonadiabatic Spin Torque}
\author{Jung-Hwan Moon$^1$, Dong-Hyun Kim$^2$, Myung Hwa Jung$^3$, and Kyung-Jin Lee$^1$$^\dagger$}
\affiliation{$^1$Department of Materials Science and Engineering, Korea University, Seoul 136-701, Korea \\
$^2$Department of Physics, Chungbuk National University, Chengju 361-763, Korea \\
$^3$Department of Physics, Sogang University, Seoul 121-742, Korea}

\date{\today}

\begin{abstract}
The current-induced resonant excitation of a magnetic vortex core is
investigated by means of analytical and micromagnetic calculations.
We find that the radius and the phase shift of the resonant motion
are not correctly described by the analytical equations because of
the dynamic distortion of a vortex core. In contrast, the initial
tilting angle of a vortex core is free from the distortion and
determined by the nonadiabaticity of the spin torque. It is
insensitive to experimentally uncontrollable current-induced
in-plane Oersted field. We propose that a time-resolved imaging of
the very initial trajectory of a core is essential to experimentally
estimate the nonadiabaticity.
\end{abstract}


\maketitle

A spin-polarized current can exert torque to a ferromagnet by
transferring spin-angular momentum, i.e. spin-transfer torque. The
spin-transfer torque provides full magnetization reversal,
steady-state precession motion, and domain wall
movement~\cite{Berger, Slon}. It is composed of adiabatic and
nonadiabatic spin torque terms in continuously varying
magnetization. The adiabatic spin torque arises from the conduction
electron spin whose projection on the film plane follows the
direction of a local magnetization, whereas the nonadiabatic torque
arises from a mismatch of the direction as a result of the momentum
transfer or the spin relaxation~\cite{Zhang, TataraPRL, Thiaville}.

Until now, the experimental threshold current density $J_C$ to
steadily move a domain wall has been reported to be about $10^8
A/cm^2$, too large for an application. In addition to the resonant
depinning~\cite{Thomas} and the use of perpendicular
anisotropy~\cite{Jung}, an increase of $\beta$, the ratio of the
nonadiabatic spin torque to the adiabatic one can reduce $J_C$
~\cite{TataraJPSJ}. Despite its importance, however, the exact value
of $\beta$ is still under debate even in theories~\cite{Barnes,
Tserkovnyak, Kohno, Xiao, Seo}; $\beta = 0$, $\beta = \alpha$, and
$\beta \neq \alpha$ where $\alpha$ is the Gilbert damping constant.
This debate is also related to which mechanism between the
Landau-Lifshitz damping and the Gilbert one is correct to describe
the energy dissipation ~\cite{Stiles}. Experimental determination of
$\beta$ is essential to resolve this debate, but experimentally
estimated values are also distributed; $\beta = 8
\alpha$~\cite{Thomas}, $\beta = \alpha$~\cite{Hayashi06}, $\beta >
\alpha$~\cite{Hayashi07}, $\beta =2\alpha$~\cite{Moriya},and $\beta
\neq \alpha$~\cite{Heyne}. Since most experiments have used the same
material (Permalloy), this wide distribution is caused by origins
irrelevant to the material itself.

The wide distribution can originate from the Joule heating, the edge
roughness of nanowire, and the in-plane component of the
current-induced Oersted field $H_{Oe}^{In}$. The Joule heating
significantly affects $J_C$ and wall velocity~\cite{Seo, Martinez}.
Therefore, it is difficult to precisely estimate $\beta$ when the
Joule heating is not negligible, i.e. $J_C
> 10^8 A/cm^2$. In a magnetic nanowire, the edge roughness distorts
the domain wall~\cite{HeJAP} and thus prevents a proper
interpretation of experimental data using theories derived for an
ideal nanowire. A way to avoid the above issues is to experimentally
study resonant motions of a magnetic vortex core (VC) in a patterned
disk by injecting an alternating current of the order of $10^7
A/cm^2$. The magnetic vortex is an ideal system for the resonant
motion study since VC can be considered as a topological point
charge which efficiently responses to external
forces~\cite{Guslienko, Kim, Moon}. It was experimentally confirmed
that the VC can be resonantly excited by an ac current~\cite{Thomas,
Moriya, Kasai, Bolte}. Even in this case, however, a very small
in-plane component of the ac current-induced alternating Oersted
field $H_{Oe}^{In}$ inhibits a precise estimation of the spin torque
parameters~\cite{Bolte, Kruger}. Note that the $H_{Oe}^{In}$ is not
a current-induced field along the thickness direction of the disk,
but an in-plane field caused by any geometrical symmetry-breaking of
the system. The driving force due to the $H_{Oe}^{In}$ of only $0.3
Oe$ is as large as $30\%$ of the total resonant
excitation~\cite{Bolte}. Such a small $H_{Oe}^{In}$ is difficult to
remove since it is caused by an uncontrollable nonuniform current
distribution due to a geometrical symmetry-breaking such as electric
contacts or notches. Therefore, an experimental way to estimate the
$\beta$ which is safe from the Joule heating and the edge roughness,
and also insensitive to the $H_{Oe}^{In}$ is highly desired.

In this letter, we propose that a direct imaging of the very initial
trajectory of VC induced by an ac current is a plausible way to
experimentally estimate $\beta$. We find that $\beta$ does not
change the resonant frequency, but affects the phase of resonant
motion. The phase-shift is $\beta$-dependent since $\beta$
determines the tilting angle of very initial trajectory measured
from the direction of the electron-flow. On the other hand, the
$H_{Oe}^{In}$ with a typical magnitude does not change the tilting
angle although it affects the steady resonant motion. More
importantly, the initial tilting angle is only one physical
observable which can be directly compared to the analytical result,
whereas the others such as the radius and the phase shift are not
correctly described by the analytical equations because of the
dynamic distortion of VC.

\begin{figure}[ttbp]
\begin{center}
\psfig{file=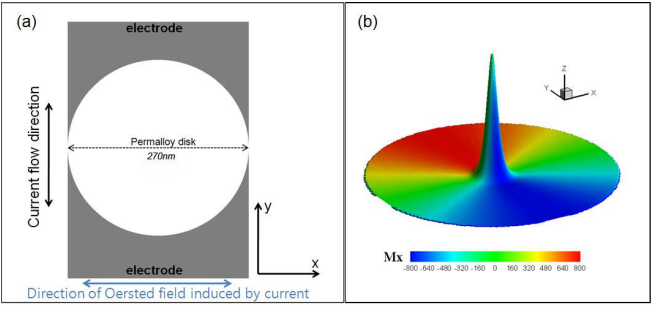,width=\columnwidth} \caption{\label{fg1} (Color
online) (a) Schematics of the model system. (b) Magnetization of a
vortex in its initial equilibrium state. The height denotes the z
component, whereas the color scale corresponds to the direction of
the x-component of magnetization.}
\end{center}
\end{figure}

The current-induced motion of VC is calculated using the Thiele's
equation with the spin-transfer torque terms (Eq.
(\ref{Thiele}))~\cite{Thiaville, Thiele, Shibata, HePRB}.
\begin{equation}\label{Thiele}
{\mathbf{G}(p) \times (\mathbf{u}-\mathbf{\dot{r}})= - {\delta
U(\mathbf{r}) \over \delta \mathbf{r}} - \alpha D \mathbf{\dot{r}} +
\beta D \mathbf{u} }
\end{equation}
where $G(p)=-G_0 p \textbf{e}_z$ is the polarity ($p\pm1$) dependent
gyrovector, $G_0$ is obtained from the spin texture as
\begin{equation}\label{G0}
{G_0 = {M_S \over \gamma} \int dV \sin(\theta) (\vec{\nabla}\theta
\times \vec{\nabla}\psi) \cdot \mathbf{e}_z, }
\end{equation}
$\theta$ ($\psi$) is the out-of-plane (in-plane) angle of the
magnetization, $M_S$ is the saturation magnetization, $\gamma$ is
the gyromagnetic ratio, $ \textbf{u} = u_0 \exp(i\omega
t)\textbf{e}_y$, $u_0 (= \mu_B J P /eM_S)$ is the amplitude of
adiabatic spin torque, $\omega$ is the angular frequency of the ac
current, $\textbf{r}(t)=X(t) +Y(t)$ is the time-dependent position
vector of VC, and $U(r)$ is the potential well. The damping tensor
$D$ is also obtained from the integration of spin texture as
\begin{equation}\label{D}
{D = -{M_S \over \gamma} \int dV [(\vec{\nabla}\theta
\vec{\nabla}\theta + \sin^2(\theta) \vec{\nabla}\psi
\vec{\nabla}\psi)]. }
\end{equation}
When VC is at the static equilibrium position, $G_0$ is $2 \pi L
M_S/\gamma$ and $D$ is $G_0 \ln(R/\delta)/2$ where $L$ is the
thickness of disk, $R$ is the vortex radius, and $\delta$ is the
core diameter.

With $X(t)=X_0\exp(i\omega t), Y(t)=Y_0\exp(i\omega t)$ and
$1+(\alpha D/G_0)^2 \sim 1$, the solutions are in the form of
$X(t)=X_1\cos(\omega t)-X_2\sin(\omega t)$ and $Y(t)=Y_1\cos(\omega
t)-Y_2\sin(\omega t)$ where
\begin{eqnarray}\label{solution}
X_1 & = & A \omega_r [
-(\omega_r^2-\omega^2)+2C^2\alpha(\beta-\alpha)\omega^2], \nonumber \\
X_2 & = & A \omega C [(\beta-\alpha)(\omega_r^2-\omega^2)+2\alpha
\omega_r^2], \nonumber \\
Y_1 & = & A \omega_r C [ (\omega_r^2-\omega^2)+2\alpha(1+C^2 \alpha
\beta)\omega^2], \nonumber \\
Y_2 & = & A \omega [(1+C^2 \alpha \beta)(\omega_r^2-\omega^2)-2C^2
\alpha \beta \omega_r^2].
\end{eqnarray}
Here, $A=u_0/[(\omega_r^2-\omega^2)^2+(2C\alpha\omega_r\omega)^2]$,
$\omega_r = \kappa/G_0$ is the resonance frequency,
$\kappa=(dU/dr)/r$ is the effective stiffness coefficient of the
potential well, and $C$ is $D/G_0 = \ln(R/\delta)/2$. From the eqs.
(\ref{solution}), one finds the radius $a(t)=\sqrt{X(t)^2+Y(t)^2}$,
and the phase shift $\phi$ between the phase of the core gyration
and that of the ac current;
\begin{eqnarray}\label{phase}
\phi & = & \tan^{-1} \Big( {X_1 \over Y_1} \Big) \nonumber \\
& = & \tan^{-1} {\Big[ {1-(\omega_r / \omega)^2 + 2C^2 \alpha (\beta
- \alpha) \over C\beta((\omega_r /
\omega)^2-1)+2C\alpha(1+C^2\alpha\beta)}\Big].}
\end{eqnarray}

To verify the validity of the analytical solutions, the
micromagnetic simulation is also performed by means of the
Landau-Lifshitz-Gilbert equation including the spin torque terms;
\begin{eqnarray}\label{LLG}
{\partial \mathbf{M} \over \partial t}= & - & \gamma \mathbf{M}
\times \mathbf{H}_{eff} + {\alpha \over M_S} \mathbf{M} \times
{\partial \mathbf{M} \over
\partial t} \nonumber \\
& + & {u_0} \nabla \mathbf{M} - {\beta u_0 \over M_s} \mathbf{M}
\times \nabla \mathbf{M}
\end{eqnarray}
where $\textbf{H}_{eff}$ is the effective field including the
external, the magnetostatic, the exchange, and the current-induced
Oersted field. The model system is a circular Permalloy disk with
the thickness of $10 nm$ and the diameter of $270 nm$ (Fig. 1 (a)).
The unit cell size is $2\times2\times10 nm^3$. The ac current with
the maximum current density of $1.25\times10^7 A/cm^2$ is applied
along the y-axis. Standard material parameters for Permalloy are
used; $M_S = 800 emu/cm^3$, $\gamma= 1.76\times10^7 Oe^{-1}s^{-1}$,
$\alpha = 0.01$, $P = 0.7$, and the exchange constant $A_{ex} =
1.3\times10^{-6} erg/cm$.

\begin{figure}[ttbp]
\begin{center}
\psfig{file=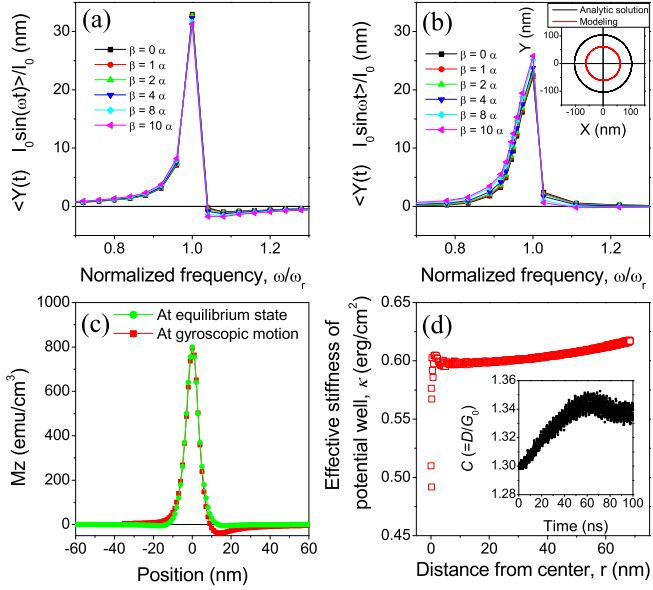,width=\columnwidth} \caption{\label{fg2} (Color
online) The  $\langle Y(t)\cdot I_0\sin(\omega t)\rangle /I_0$ as a
function of the frequency obtained from (a) analytic solution of
Thiele's equation and (b) micromagnetic simulation. (c) Comparison
of the shape of vortex core, and (d) variation of $\kappa$ as a
function of the gyration radius. The inset of (b) shows the vortex
core trajectories. The inset of (d) shows variation of the parameter
$C$ ($=D/G0$) with the time evolution. Both insets are obtained at
$\beta =0$ and $\omega=\omega_r$.}
\end{center}
\end{figure}

First, we assume $H_{Oe}^{In}=0$ in order to investigate the effect
originating exclusively from $\beta$ on the resonant motion. We will
recall the effect of $H_{Oe}^{In}$ in the last part. In Fig. 2 (a)
and (b), we show analytical and modeling results of the time
averaged value $\langle Y(t)\cdot I_0\sin(\omega t)\rangle /I_0$ at
various $\beta$-terms. To obtain analytical results, we use $C=1.3$
because $R$ is $135 nm$ and $\delta$ is $10 nm$, determined from the
micromagnetic configuration at the initial equilibrium state.
$\langle Y(t)\cdot I_0\sin(\omega t)\rangle /I_0$ shows a peak at
the resonance frequency $\omega_r$ of $360 MHz$. $\omega_r$ does not
change with $\beta$ whereas the peak structure becomes more
asymmetric as $\beta$ increases. In spite of qualitative agreement,
however, analytical results are quantitatively different from
modeling ones. This is because the radius of gyroscopic motion is
different between the modeling result and the analytic solution
(inset of Fig. 2(b)). We attribute this difference to a dynamic
distortion of VC. As shown in Fig. 2(c), the VC shape in the initial
equilibrium state is symmetric whereas it in a dynamic motion is
asymmetric. The distortion changes the gyroscopic parameter $G_0$,
the damping tensor $D$ (thus, the parameter $C$) and the effective
stiffness coefficient $\kappa$ since all parameters are determined
by details of the spin texture (See, eqs. (\ref{G0}), (\ref{D}), and
the definition of $\kappa$). From micromagnetic spin textures of the
vortex in the dynamic motion, we find that both $C$ and $\kappa$
increase from the initial equilibrium values because of the dynamic
distortion (Fig. 2(d)). The increase of $\kappa$ is much larger than
that of $C$, and responsible for the reduced radius in the modeling
results. This increase of $\kappa$ occurs at the very initial time
stage, indicating that the assumption of the rigid VC and potential
well is invalid except for the very initial trajectory. Therefore,
even when $H_{Oe}^{In}$ is zero, it may be difficult to deduce
important parameters such as $P$ by directly comparing analytical
solutions with experimental measurements of the steady gyroscopic
motion.

Nevertheless, it is worthwhile investigating how $\beta$ induces the
asymmetry in the peak structure. Fig. 3(a) shows analytical phase
shifts at various $\beta$-terms. As $\beta$ increases, the absolute
value of the phase shift $\phi$ decreases (increases) for the
frequency smaller (higher) than $\omega_r$. In other words, a
vertical offset $\phi(\beta)-\phi(\beta=\alpha)$ of the phase shift
increases with increasing $\beta$. From the eq. (\ref{solution})
with $\alpha \ll
 1$ and $\beta \ll 1$, one can find that the vertical offset is approximately
$\tan^{-1}[C(\beta-\alpha)]$ and thus dependent on $\beta$. This is
why the peak structure becomes more asymmetric as increases.
However, quantitative disagreement between analytic solution and
modeling result was again observed (inset of Fig. 3(a)). The
difference becomes larger as the radius of core gyration increases,
i.e. the frequency approaches $\omega_r$. This is also caused by the
dynamic distortion as explained above.

\begin{figure}[ttbp]
\begin{center}
\psfig{file=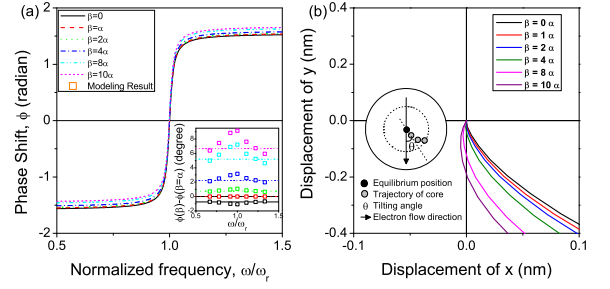,width=\columnwidth} \caption{\label{fg3} (Color
online) (a) Phase shift as a function of the frequency for various
$\beta$-terms. (b) Initial vortex core trajectories for various
$\beta$-terms ($\omega=\omega_r$). Inset of (a) shows the vertical
offset of the phase shift for various $\beta$-terms. In the inset,
open symbols are obtained from micromagnetic modeling. Inset of (b)
describes the definition of the initial tilting angle.}
\end{center}
\end{figure}

The vertical offset is $\beta$-dependent since the initial tilting
angle $\theta_{int}$ is determined by $\beta$ (Fig. 3(b)). VC
initially moves along the direction of the electron-flow. Because of
the imbalance of magnetostatic field, VC experiences the centripetal
force and starts to undergo a gyration motion. In the absence of
$H_{Oe}^{In}$, $\theta_{int}$ of the initial trajectory can be
obtained from the eq. (\ref{Thiele}) by dropping the potential
gradient term since VC is initially at the bottom of the potential
well where the gradient is zero. When $H_{Oe}^{In}$ is nonzero, the
potential gradient is no longer zero and could affect
$\theta_{int}$. In order to investigate the effect of $H_{Oe}^{In}$
on $\theta_{int}$, we perform micromagnetic simulations for the
initial trajectory with and without taking into account
$H_{Oe}^{In}$. We assume that an alternating $H_{Oe}^{In}$ is
applied along the x-axis and its magnitude is $0.3 Oe$ which is
similar with the estimated value in the experiment of the
Ref.~\cite{Bolte}. As shown in Fig. 4(a), the effect of
$H_{Oe}^{In}$ on the very initial trajectory is negligible whereas
the difference in trajectories between the two cases becomes larger
and larger as the time evolves. This insensitivity of the initial
trajectory and thus $\theta_{int}$ to $H_{Oe}^{In}$ is valid for a
different $\beta$ (not shown). Thus, we drop the potential gradient
term in the eq. (\ref{Thiele}) to derive $\theta_{int}$. With $p=+1$
and the direction of the initial current along the +y-axis, one
finds
\begin{eqnarray}\label{init1}
\alpha D \dot{X} - G_0 \dot{Y} & = & G_0 u_0, \nonumber \\
G_0 \dot{X} + \alpha D \dot{Y} & = & - \beta D u_0
\end{eqnarray}
where $\dot{X}$ and $\dot{Y}$ are the velocity along the x- and
y-axis, respectively. By solving the eqs. (\ref{init1}) for
$\dot{X}$ and $\dot{Y}$,
\begin{eqnarray}\label{init2}
\dot{X} & = & G_0 u_0 D {\alpha - \beta \over \alpha^2 D^2 + G_0^2}, \nonumber \\
\dot{Y} & = & -u_0 {G_0^2+\alpha\beta D^2 \over \alpha^2 D^2 +
G_0^2}.
\end{eqnarray}
The initial tilting angle $\theta_{int}$ is given by
\begin{equation}\label{angle}
\theta_{int} = \tan^{-1} \Big( {\dot{X} \over -\dot{Y}} \Big) =
\tan^{-1} \Big( {C(\alpha - \beta) \over 1+\alpha\beta C^2} \Big).
\end{equation}
Note that the eq. (\ref{angle}) is equivalent to the equation for
the vertical phase-shift with considering the sign of the initial ac
current and $\alpha\beta \ll 1$. It confirms that the vertical shift
originates from the $\beta$-dependent $\theta_{int}$. Fig. 4(b)
shows $\theta_{int}$ as a function of $\beta/\alpha$ for various
values of $C$. It should be noted that $\theta_{int}$'s obtained
from modeling are in excellent agreement with analytical ones in
contrast to the radius and the phase shift. It is because VC retains
its equilibrium shape at the very initial time stage where $C$ and
$\kappa$ hardly changes. For the tested Permalloy disk ($C = 1.3$),
the difference in $\theta_{int}$ between $\beta= 0$ and $\beta=
8\alpha$ is about $6$ degree which may be small to experimentally
measure. However, the $\beta$-dependent $\theta_{int}$ becomes
larger as the parameter $C$ increases. For instance, the difference
in $\theta_{int}$ between $\beta= 0$ and $\beta= 8\alpha$ is about
$18$ degree at $C = 4$. The $C$ increases as the disk diameter (core
diameter) increases (decreases). The core diameter $\delta$
decreases with decreasing the disk thickness~\cite{Altbir}. The
equilibrium value of $C$ can be determined by micromagnetic
calculation or MFM imaging for a given disk geometry. Consequently,
a time-resolved magnetic imaging with high spatial resolution such
as X-ray microscopy~\cite{Fischer} for observation of the very
initial trajectory of VC in a wider and thinner disk is a possible
way to experimentally estimate $\beta$.

\begin{figure}[ttbp]
\begin{center}
\psfig{file=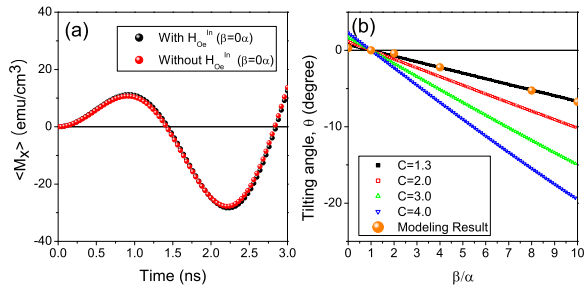,width=\columnwidth} \caption{\label{fg4} (Color
online) (a) Effect of in-plane Oersted field ($H_{Oe}^{In}$) on the
initial trajectory of vortex core ($\beta =0$ and
$\omega=\omega_r$), and (b) effect of $\beta$ and $C$ on the initial
tilting angle.}
\end{center}
\end{figure}

 This work was supported by a Korea Research Foundation Grant
funded by the Korean Government (MOEHRD) (KRF-2006-311-D00102) and
the Korea Science and Engineering Foundation (KOSEF) through the
National Research Laboratory Program funded by the Ministry of
Science and Technology (No. M10600000198-06J0000-19810).

($\dagger$) Corresponding email: kj\_lee@korea.ac.kr



\end{document}